

\magnification=\magstep1
\centerline{\bf DETERMINATION OF ${\bf f(\infty)}$ FROM THE ASYMPTOTIC}
\centerline{\bf SERIES FOR ${\bf f(x)}$ ABOUT ${\bf x=0}$}
\bigskip
\bigskip
\bigskip
\centerline{Carl M. Bender and Stefan Boettcher\footnote{*}{Current
Address: Brookhaven National Laboratory, Upton, NY 11973}}
\medskip
\centerline{Department of Physics}
\medskip
\centerline{Washington University}
\medskip
\centerline{St. Louis, MO 63130}
\bigskip
\bigskip
\bigskip
\bigskip
\bigskip
\centerline{\bf ABSTRACT}
\bigskip
A difficult and long-standing problem in mathematical physics concerns the
determination of the value of $f(\infty)$ from the asymptotic series for $f(x)$
about $x\!=\!0$. In the past the approach has been to convert the asymptotic
series to a sequence of Pad\'e approximants $\{P^n_n(x)\}$ and then to evaluate
these approximants at $x\!=\!\infty$. Unfortunately, for most physical
applications the sequence $\{P^n_n(\infty)\}$ is slowly converging and does not
usually give very accurate results. In this paper we report the results of
extensive numerical studies for a large class of functions $f(x)$ associated
with strong-coupling lattice approximations. We conjecture that for large $n$,
$P^n_n(\infty)\!\sim\!f(\infty)+B/\ln n $. A numerical fit to this asymptotic
behavior gives an accurate extrapolation to the value of $f(\infty )$.
\footnote{}{PACS numbers: 02.60.+y, 11.15.Me, 02.90.+p}
\footnote{}{hep-lat/9312031}
\vfill \eject

There are many examples in mathematical physics where it is necessary to
calculate the value of a function $f(x)$ at $x\!=\!\infty$ but where $f(x)$ can
only be determined perturbatively for small $x$.$^{1,2,3}$ The result of such a
perturbative calculation yields the first few terms of the asymptotic series
for
$f(x)$:

$$f(x)~=~x^{\alpha}~\sum^{\infty}_{n=0}~b_n x^n~~.\eqno(1)$$
The problem is then to use this series to determine $f(\infty)$
assuming, of course, that $f(\infty)$ exists.

A natural approach$^{4,5,6,7,8}$ to solving this problem is to raise both sides
of (1) to the power $1/\alpha$,
$$f(x)^{1/\alpha}~=~ x~\sum_{n=0}^{\infty}~c_nx^n~~, \eqno(2)$$
and then to convert the right side of (2) to a finite sequence of Pad\'e
approximants, $P^n_n(x)$, $n=1,2,3,\ldots,N$, where $2N-1$ is the highest order
of the perturbative calculation used to obtain (1). Since $P^n_n(\infty)$
exists, we compute the sequence of Pad\'e extrapolants $[P^n_n(\infty)]^
{\alpha}$ with the hope that as $n$ increases, this sequence rapidly approaches
$f(\infty)$.

While this extrapolation procedure works well in some cases one encounters an
important class of functions for which this procedure is rather ineffective.
For
this class of functions the coefficients $b_n$ in (1) have two characteristic
properties:

($i$) $b_n \sim n! $ as $n\to\infty$;

($ii$) the $b_n$ exhibit a doubly-alternating sign pattern
$+,~+,~-,~-,~+,~+,~-,~-,~\ldots~~$.

\noindent
Asymptotic series whose coefficients exhibit properties ($i$) and ($ii$) arise
in strong-coupling lattice expansions in quantum field theory.$^{4,5,6,7,8}$
The
problem of determining the value of $f(\infty)$ corresponds to extrapolating to
the continuum limit of the lattice theory. For such functions, experience shows
that while $[P^n_n(\infty)]^{\alpha}$ often comes close to $f(\infty)$, the
sequence of Pad\'e extrapolants appears to level off at some value that differs
from $f(\infty)$ by several percent. This raises the question of whether the
sequence actually converges to $f(\infty)$ or to some other limit near
$f(\infty)$.

Here is a simple example that illustrates this problem: Consider the series
$$f(x)~=~\sum_{n=0}^{\infty}~a_n x^n~~,\eqno(3)$$
where
$$a_{2n}~=~(-1)^n~(2n)!~~,~~~~a_{2n+1}~=~(-1)^n~2~(2n)!~~.\eqno(4)$$
We can evaluate this series exactly using Borel summation. The result is the
integral representation
$$f(x)~=~(1+2x)~\int\limits_0^{\infty}dt~{{e^{-t}}\over{1+x^2
t^2}}~~,\eqno(5)$$
from which we determine that
$$f(\infty)~=~\pi~~.\eqno(6)$$

Let us try to obtain this result numerically by converting the series in (3) to
a sequence of Pad\'e extrapolants. On Fig.~1 we plot the value of
$P^n_n(\infty)$ as a function of $1/n$ for $n=1,~2,~3,\ldots,~20$. Observe that
the
Pad\'e approximants form a zigzag sequence$^{9}$ that seems to tend to a
limiting value on the y-axis (corresponding to $n\!=\!\infty$). If we
extrapolate the sequence to $n\!=\!\infty$ using a straight line, the sequence
appears to have a limit that is too small (roughly 10\% smaller than $\pi$).
However, if we include many more terms ($n\!=1,2,3,\ldots,300$) in the sequence
(see Fig. 2), we can see that the sequence actually lies on a curve that
becomes
steeper as $n\!\to\!\infty$. Figure 2 indicates that the asymptotic behavior of
$P^n_n(\infty)$ for large $n$ cannot have a straight-line form $\pi+B/n$.

Because the sequence in Fig.~2 lies on a curve and not on a straight line it is
difficult to predict its limiting value. A detailed numerical fit of the form
$$ P^n_n(\infty)~\sim~\pi + {B \over {n^C}}$$
does not work for any value of $C$. However, a more general numerical fit of
the
form
$$ P^n_n(\infty)~\sim~\pi + {B \over {n^C (\ln n)^D}}$$
gives $C\!=\!0\pm 0.1$ and $D\!=\!1\pm 0.1$. Thus, we believe that
$$ P^n_n(\infty)~\sim~f(\infty) + {B \over {\ln n}}~~.  \eqno(7)$$
In fact, a fit of the form $P^n_n(\infty)\!=\!A+B/\ln n$ gives $A\!=\!\pi$ with
a relative error of less than 1\%, which is an order-of-magnitude improvement
over the result of extrapolating the curve in Fig.~1. In Fig.~3 we plot the
Pad\'e extrapolants shown on Fig.~2 as a function of $1/\ln (20n)$.$^{10}$
Observe that the zigzag curve now lies along a straight line that appears to
approach the value $\pi$.

Next, we consider the more complicated case of the Stirling series expansion
for
the factorial function:
$$\eqalign{
f(x)~\sim~ x^{-{1 \over 2}}~[ 1 &+ {1 \over {12}} x + {1 \over {288}} x^2
- {{139} \over {51,840}} x^3 - {{571} \over {2,488,320}} x^4 \cr
&+{{163,879} \over {209,018,880}} x^5 + {{5,246,819} \over {75,246,796,800}}
x^6
- \ldots ]~~~(x\to 0)~~, \cr} \eqno(8)$$
where
$$f(x)~=~{1 \over {\sqrt{2 \pi}}}~e^{1 \over x}~x^{1 \over x}~({1 \over x})!~~.
\eqno(9)$$
Note that the series in (8) exhibits properties ($i$) and ($ii$).$^{11}$

Since $0!\!=\!1$ we see that
$$ f(\infty)~=~{1 \over {\sqrt{2 \pi}}}~=~0.398942 \ldots ~~. \eqno(10)$$
We attempt to calculate $f(\infty)$ by raising (8) to the power $-2$, computing
$P^n_n(x)$, and forming the sequence of extrapolants $[P^n_n(\infty)]^{-1/2}$.
As in the previous example if we plot these extrapolants as a function of $1/n$
for $n=1,~2,~3,\ldots,~90$ (see Fig.~4) we see that the extrapolants do not lie
along a straight line. From just the first 20 extrapolants we would predict the
value of $f(\infty)$ to be about 0.42, which is 5\% high. However, it is clear
that the extrapolants follow a curve which becomes increasingly steep as $n\to
\infty$.
If we replot the data as a function of $1/\ln (20n)$ (see Fig.~5) we
see that the same extrapolants lie on a straight line that appears to intersect
the $y$-axis at precisely the correct result in (10).$^{10}$ A numerical fit
of the form in (7) gives $f(\infty)$ correct to about 1\%.

Our third example is taken from a quantum field theory calculation. Consider
a $g\phi^{2K}$ self-interacting scalar quantum field theory in $D$-dimensional
Euclidean space. In recent papers$^{12,13}$ we showed how to express the free
energy for such a theory as a series in powers of $D$ in the limit of
strong bare coupling, $g\to\infty$. The coefficient of $D$ in
this dimensional expansion can be expressed as the logarithm of an infinite
series in powers of $x$ like that in (1); we denote this infinite series
$f(x)$.$^{14}$ For all $K$ the coefficients of the series $f(x)$ grow like $n!$
and have a doubly-alternating sign pattern. The expansion parameter, $x=[ga^{2K
-D(K-1)}]^{-1/K}$, is a dimensionless combination of the coupling constant $g$
and the lattice spacing $a$, which is small for fixed $a$ and large
$g$. But in the continuum limit for which $a\to 0$, we have
$x\to\infty$. Thus, we are confronted with the problem addressed in this paper,
namely that of computing $f(\infty)$.

As a special case, we consider the free theory corresponding to $K\!=\!1$. For
this case there is a closed-form expression for $f(x)$:
$$f(x)~=~x~\exp\left\{2\int_0^{\infty}dt\,e^{-t}\,\ln
[e^{-xt}I_0(xt)]\right\}~,.\eqno(11)$$
where $I_{0}$ is the modified Bessel function of order $0$.
In the continuum limit, $x\to\infty$, we have
$$f(\infty)~=~{e^{\gamma}\over{2\pi}} ~=~0.28347\ldots ~~.\eqno(12)$$
The first few terms of the asymptotic series representation for $f(x)$ in
powers
of $x$ are
$$\eqalign{f(x&)~=~x\Big(1-2x+3x^2-{{10}\over{3}}x^3+{{29}\over{12}}x^4
-{{11}\over{10}}x^5+{{391}\over{180}}x^6 -{{2389}\over{630}}x^7
-{{5303}\over{448}}x^8  \cr
&+{{2602051}\over{90720}}x^9+{{159662191}\over{907200}}x^{10}
-{{651255947}\over{1663200}}x^{11}-{{435388434359}\over{119750400}}x^{12}+\ldots \Big)~~.\cr}\eqno(13)$$
If we convert the first 250 terms in this series to a sequence of Pad\'e
extrapolants, we obtain results similar to those of the two examples above. On
Fig.~6 we plot this sequence of extrapolants as a function of $1/n$ and observe
that the extrapolants lie along a curve that is difficult to extrapolate to its
value at $n\!=\!\infty$. However, on Fig.~7 we plot these same extrapolants
versus $1/\ln (4n)$ and obtain a straight line whose limit is clearly very
close
to the exact answer given in (12).$^{15}$

On the basis of our numerical studies of these examples we conjecture that the
asymptotic behavior in (7) of the $n$th Pad\'e extrapolant is generic if the
coefficients exhibit properties ($i$) and ($ii$). We may express the Pad\'e
extrapolant $P_n^n (\infty)$ constructed from the asymptotic series
$x\sum_{n=0}
^\infty c_n x^n$ as a ratio of two determinants, an $(n+1) \times (n+1)$
determinant divided by an $n \times n$ determinant.$^{16}$ Thus, a simple way
to
state this conjecture is in terms of determinants of large-dimensional
matrices:
$$ P_n^n (\infty)~=~{{\left|\matrix{
0&c_0&c_1&c_2&\cdots&c_{n-1}\cr
c_0&c_1&c_2&c_3&\cdots&\vdots\cr
c_1&c_2&c_3&c_4&\cdots&\vdots\cr
c_2&c_3&c_4&c_5&\cdots&\vdots\cr
\vdots&\vdots&\vdots&\vdots&\ddots&\vdots\cr
c_{n-1}&\cdot&\cdot&\cdot&\cdots&c_{2n-1}\cr
}\right|} \over
{\left|\matrix{
c_1&c_2&c_3&c_4&\cdots&c_n\cr
c_2&c_3&c_4&c_5&\cdots&\vdots\cr
c_3&c_4&c_5&c_6&\cdots&\vdots\cr
c_4&c_5&c_6&c_7&\cdots&\vdots\cr
\vdots&\vdots&\vdots&\vdots&\ddots&\vdots\cr
c_n&\cdot&\cdot&\cdot&\cdots&c_{2n-1}\cr}\right|}}
{}~\sim~f(\infty)~+~{B\over{\ln n}}~~.  \eqno(14)$$
\bigskip
\bigskip

We are indebted to L. N. Lipatov, G. McCartor, and M. VanDyke for useful
discussions. We thank the U. S. Department of Energy for funding this research
and one of us, SB, thanks the U. S. Department of Education for financial
support in the form of a National Need Fellowship.
\vfill \eject

\centerline{\bf REFERENCES}
\bigskip
\item{$^1$} G. Parisi, Phys. Lett. {\bf 69B}, 329 (1977).
\medskip
\item{$^2$} In the field of fluid mechanics, M. VanDyke [J. Fluid Mech.
{\bf 44}, 813 (1970)] used a 24-term series in powers of the Reynolds number
$R$ to obtain the drag coefficient of a sphere in the Oseen linearization.
Using an Euler transformation he was able to extrapolate the series to its
value at $R = \infty$. The accuracy of his result was about 5\%.
VanDyke applied similar techniques to a 24-term series in powers of Dean number
$K$ for laminar flow through a loosely coiled pipe [J. Fluid Mech. {\bf 86},
129 (1978)]. Extrapolating the series to its value at $K=\infty$, VanDyke
obtained the controversial result that there is a 1/20-power growth of the
friction factor. VanDyke's series differ from the series considered in this
paper in that his series have a finite (nonzero) radii of convergence.
\medskip
\item{$^3$} C. M. Bender, F. Cooper, G. S. Guralnik, and D. H. Sharp, Phys.
Rev.
D {\bf 19}, 1865 (1979).
\medskip
\item{$^4$} C. M. Bender, F. Cooper, G. S. Guralnik, R. Roskies, and D. H.
Sharp, Phys. Rev. Lett. {\bf 43}, 537 (1979).
\medskip
\item{$^5$} C. M. Bender, Los Alamos Science {\bf 2}, 76 (1981).
\medskip
\item{$^6$} C. M. Bender and D. H. Sharp, Phys. Rev. D {\bf 24}, 1691 (1981).
\medskip
\item{$^7$} R. J. Rivers, Phys. Rev. D {\bf 20}, 3425 (1979).
\medskip
\item{$^8$} R. J. Rivers, Phys. Rev. D {\bf 22}, 3135 (1980).
\medskip
\item{$^9$} The zigzag oscillation is connected with the doubly-alternating
sign
pattern of the coefficients $a_n$.
\medskip
\item{$^{10}$} The scale factor 20 in the argument of the logarithm function
tends to give a particularly smooth fit to the data. By choosing the scale
factor appropriately we can minimize the effect of the next term in the fit,
which presumably has the form $(\ln n)^{-2}$.
\medskip
\item{$^{11}$} {\it Advanced Mathematical Methods for Scientists and
Engineers},
C. M. Bender and S. A. Orszag (McGraw-Hill, New York, 1978), pp. 218-227.
\medskip
\item{$^{12}$} C. M. Bender, S. Boettcher, and L. N. Lipatov, Phys. Rev. Lett.
{\bf 65}, 3674 (1992).
\medskip
\item{$^{13}$} C. M. Bender, S. Boettcher, and L. N. Lipatov, Phys. Rev. D {\bf
46}, 5557 (1992).
\medskip
\item{$^{14}$} Coefficients of higher powers of $D$ can also be expressed in
terms of infinite series in powers of $x$.
\medskip
\item{$^{15}$} For $K\!>\!1$ the form of the series and its extrapolants are
similar to the $K\!=\!1$ case. However, when $K\!>\!1$ the quantum field theory
is not free and it is difficult to calculate more than half a dozen
extrapolants.
\medskip
\item{$^{16}$} See Ref. 11, pp. 383-384.

\vfill \eject
\centerline{\bf FIGURE CAPTIONS}
\bigskip

\noindent
Figure 1. The first twenty Pad\'e extrapolants $P^n_n(\infty)$ for the series
whose coefficients are given in (4). The approximants are plotted as a function
of $1/n$ such that the vertical axis corresponds to $n\!=\!\infty$. Successive
Pad\'e extrapolants are connected by line segments to form a zigzag line. The
vertices of the zigzag line are the values of the approximants. One would think
that it is easy to extrapolate by eye to the limit of the sequence (the point
where it intersects the vertical axis). However, the zigzag line of
extrapolants
appears destined to intersect the vertical axis well below the exact answer,
$\pi$, which is indicated by the horizontal line.
\medskip

\noindent
Figure 2. Same as in Fig. 1 except that we have plotted the first 300
approximants. Having plotted many more approximants than in Fig.~1, we
believe that the zigzag line of approximants is curving upward; it could
well intersect the vertical axis at $\pi$. It is difficult to guess where such
a curve might cross the vertical axis.
\medskip

\noindent
Figure 3. The same approximants as in Fig. 2 now plotted versus $1/ \ln (20n)$.
The zigzag line of approximants seems to be tending to $\pi$, the correct value
of $f(\infty)$.
\medskip

\noindent
Figure 4. The first 90 Pad\'e extrapolants constructed from the Stirling series
(8) for the factorial function. As in Fig.~2, the zigzag line of extrapolants,
when plotted versus $1/n$, begins to curve as $n$ gets large. For small values
of $n$ the zigzag line appears to be headed for an intersection with the
vertical axis that is too large compared with the correct answer $1/\sqrt
{2\pi}
\!=\! 0.398942\ldots~~$. For larger values of $n$, the curvature of the zigzag
line makes it too difficult to predict where the line will cross the vertical
axis.
\medskip

\noindent
Figure 5. The same approximants as in Fig.~4 now plotted versus $1/ \ln (20n)$.
As in Fig.~3, the zigzag line of approximants seems to be tending to the
correct
value of $f(\infty)$, which in this case is $1/\sqrt{2\pi}\!=\!0.398942$.
Apparently, one can
use the Stirling series for $(1/x)!$, which is valid as $x\to 0$, to
calculate $0!\!=\!1$.
\medskip

\noindent
Figure 6. The first 125 Pad\'e extrapolants constructed from the asymptotic
series (13) taken from a field-theoretic calculation. As in Figs.~2 and 4, the
zigzag line of extrapolants, when plotted versus $1/n$, begins to curve as $n$
gets large. For small values of $n$ the zigzag line appears to be headed for an
intersection with the vertical axis that is too large compared with the correct
answer $e^{\gamma}/(2\pi)\!=\! 0.28347\ldots~$. For larger values of $n$, the
curvature of the zigzag line makes it too difficult to predict where the line
crosses the vertical axis.
\medskip

\noindent
Figure 7. The same approximants as in Fig.~6 now plotted versus $1/\ln (4n)$.
As in Figs.~3 and 5, the zigzag line of approximants seems to be tending to
the correct value of $f(\infty)$, which in this case is $3.52775\ldots~~$.
\bye